# High Resolution Magnetic Resonance Spectroscopy Using Solid-State Spins

Dominik B. Bucher,[1,2]* David R. Glenn,[2]* Junghyun Lee,[3] Mikhail D. Lukin,[2] Hongkun Park,[2,4] Ronald L. Walsworth[1,2]#

1. Harvard-Smithsonian Center for Astrophysics, Cambridge, MA
2. Department of Physics, Harvard University, Cambridge, MA
3. Department of Physics, Massachusetts Institute of Technology, Cambridge MA
4. Department of Chemistry and Chemical Biology, Harvard University, Cambridge, MA

* These authors contributed equally to this work.
# Corresponding author.  Email: rwalsworth@cfa.harvard.edu

## Abstract:

**We demonstrate a synchronized readout (SR) technique for spectrally selective detection of oscillating magnetic fields with sub-millihertz resolution, using coherent manipulation of solid state spins. The SR technique is implemented in a sensitive magnetometer (~50 picotesla/Hz$^{1/2}$) based on nitrogen vacancy (NV) centers in diamond, and used to detect nuclear magnetic resonance (NMR) signals from liquid-state samples. We obtain NMR spectral resolution ~3 Hz, which is nearly two orders of magnitude narrower than previously demonstrated with NV based techniques, using a sample volume of ~1 picoliter. This is the first application of NV-detected NMR to sense Boltzmann-polarized nuclear spin magnetization, and the first to observe chemical shifts and J-couplings.**

## Introduction:

Nuclear magnetic resonance (NMR) spectroscopy is an important analytical tool in modern chemistry, structural biology, and materials research. Conventional NMR relies on inductive detection and requires sample volumes of 0.1 – 1 mL [1], although alternative detection technologies including microcoils [2], superconducting quantum interference devices (SQUIDs) [3], and atomic magnetometers [4] have been demonstrated to improve sensitivity or allow reduced sample volumes. Recently, it was shown that nitrogen vacancy (NV) centers in diamond can be used to detect NMR from nanoscale sample volumes [5,6], with sufficient sensitivity to detect the time-varying magnetic field produced by a single protein [7]. However, the best reported spectral resolution for NV-based NMR detection, achieved using correlation spectroscopy [8], is $\Delta f$ ~ 210 Hz [9]. This is considerably broader than the < 10 Hz resolution needed to observe J-couplings or chemical shifts at typical static magnetic fields ($B_0$ ~ 0.1 T) used in NV experiments, making existing NV-based techniques unsuitable for many practical molecular NMR applications.

To date, two key challenges have limited the spectral resolution of NV-detected NMR techniques. First, the interrogation duration was set by the spin state lifetime of the NV ($T_1 \sim 3$ ms), or of a proximal solid-state nuclear spin used to create a quantum memory ($T_1 \sim 5 - 50$ ms) [9,10], both of which are orders of magnitude shorter than typical coherence times of nuclear spins in liquid samples ($T_2 \sim 1$ s). Second, for NV-NMR with nanoscale sample volumes, the thermal spin polarization at room temperature and $B_0 \sim 0.1$ T is small ($\sim 3 \times 10^{-7}$) compared to statistical fluctuations in the spin polarization (which scale as $1/N^{\frac{1}{2}}$, where $N$ is the number of spins in the volume) [11]. All nanoscale NV-NMR experiments have therefore measured the spin-noise-induced variance in the local magnetic field, which is large enough to detect easily ($\sigma_B \approx 0.6$ µT RMS for NV depth $d_{NV} = 5$ nm below the diamond surface and a sample of pure water [12]), but has a short correlation time limited by diffusion of sample spins through the sensing volume. For example, using $(\pi \tau_c)^{-1} \approx \frac{6}{\pi} D V^{2/3}$, where $\tau_c$ is the noise correlation time, $D$ is the bulk diffusion coefficient, and $V \approx (5$ nm$)^3$ is the sensing volume, we find $(\pi \tau_c)^{-1} \approx 150$ MHz for water ($D_w = 2 \times 10^{-9}$ m$^2$s$^{-1}$), or $(\pi \tau_c)^{-1} \approx 25$ kHz for viscous oil ($D_{oil} = 0.3 \times 10^{-9}$ m$^2$ s$^{-1}$). In the present study, we address these challenges of NV-NMR detection via (i) a synchronized readout pulse sequence to allow coherent interrogation of sample nuclei over many NV measurements [13], and (ii) implementation of this experimental protocol in an NV ensemble instrument with sufficient sensitivity to measure the thermal spin polarization instead of statistical fluctuations.

## Results:

We employ a synchronized readout (SR) technique to coherently measure oscillating magnetic signals, e.g., the free induction decay (FID) in liquid-state NMR, for a duration greatly exceeding the $T_2$ coherence time of the NV. The SR protocol (Fig. 1) consists of a series of concatenated NV magnetometry sub-sequences, interspersed with projective NV spin state readouts, all synchronized to an external clock [13]. The protocol is defined with respect to a particular central frequency $f_0$. The magnetometry sub-sequences are all identical, each consisting of an initial π/2 pulse, followed by a train of π pulses applied at a rate of 2 $f_0$, and ending with another π/2 pulse. The initial and final π/2 pulses are chosen to be 90° out of phase, such that the final NV spin population is linearly dependent on the amplitude of the oscillating magnetic field signal. (This is different from previous NV-detected NMR protocols, where magnetometry pulse sequences yielded a quadratic dependence of the final NV state on the magnetic field signal in order to sense statistical fluctuations in the sample magnetization.) At every SR readout step, the accumulated NV spin population is measured via spin-state-dependent fluorescence, and the NV spin is repolarized.

The delay between the start of successive magnetometry sub-sequences in the SR protocol is an integer number of periods at the central frequency, $\tau_{SR} = k / f_0$. Thus, a magnetic signal at $f_0$ produces exactly the same NV phase accumulation during each repetition of the sub-sequence, and the mean NV fluorescence intensity is the same at each readout step. By contrast, a magnetic signal that is slightly detuned from the central frequency, $f = f_0 + \delta f$, is advanced in phase at the start of each SR iteration by $\delta \phi = \tau_{SR} \delta f$. In this case, the mean NV fluorescence changes at every readout, and the time series of readouts will oscillate at frequency $\delta f$. In effect, the SR protocol mixes down the magnetic signal by $f_0$. The frequency resolution of the fluorescence time series data is set by the total duration of the SR protocol and is therefore limited, in principle, only by the stability of the external clock. The range of signal frequencies $f$ that can be detected without aliasing is given by $1/(2\tau_{SR})$. We note that an important condition for the successful application of the SR technique, which is not necessarily satisfied in other NV-detected NMR protocols, is that the NV must be weakly coupled to the magnetic signal source to avoid line broadening due to back-action at each SR measurement iteration.

We first applied the SR technique to a magnetic sensor consisting of a single NV center, imaged in a confocal microscope (Fig. 2a). A nearby coil antenna was excited continuously (without gating or triggering of the sources) to produce a magnetic signal consisting of three closely-spaced frequencies around 3.7325 MHz; and the SR protocol was performed at a central frequency $f_0$ = 3.7313 MHz (period 268 ns). Magnetometry was carried out using an CPMG-32 dynamical decoupling sequence [14]; the SR cycle period was $\tau_{SR}$ = 75.04 μs; and the total experiment duration was $T = n\ \tau_{SR}$ = 112.5 s, for $n = 1.5 \times 10^6$ the number of readouts. We tuned the signal strength such that magnetic field amplitude at the NV center was ~3 μT, corresponding to the maximum fluorescence contrast for a single CPMG-32 sequence. Due to finite optical collection efficiency, each SR fluorescence readout detected a mean of only ~0.03 photons. We therefore repeated the SR protocol 100 times, and constructed a periodogram from the full data set. In the resulting power spectrum, the three signal peaks were clearly distinguishable, and the spectral resolution was 5.2 mHz (FWHM).

To improve sensitivity and eliminate the need for signal averaging, we repeated the same measurement using an NV ensemble magnetic sensor that integrates fluorescence from a total of ~5 × 10$^8$ NV centers. We again observed a spectral resolution of 5.2 mHz (FWHM), this time in a single SR experiment without averaging (Fig. 2b). We then extended the SR protocol duration to T = 3 × 10$^3$ s, using $\tau_{SR}$ = 1.2 ms and $n$ = 2.5 × 10$^6$, and recorded linewidths of 0.4 mHz (Fig. 2c), again without averaging. This spectral resolution was approximately 5 orders of magnitude narrower than previously demonstrated for non-SR magnetic signal detection using NV centers, and was likely limited by pulse timing jitter and/or oscillator phase noise in the waveform generator used to synthesize the SR magnetometry sub-sequences.

Detection of magnetic resonance signals with high spectral resolution using the SR technique requires long sample coherence times. Because molecular diffusion necessarily limits the correlation times of magnetic noise in nanoscale samples, we choose to operate with larger volumes such that the mean thermal magnetization, $M_z$, is greater than the distribution width of statistical fluctuations in the magnetization, $\sigma_M$. The sample temperature and the magnitude of the static magnetic field $B_0$ determine the thermal polarization, and hence the minimum detection volume. For experimental convenience in driving magnetometry pulse sequences at the NV Larmor frequency, we typically operate at $B_0$ = 88 millitesla. Then, taking protons in water as a characteristic sample, the condition $M_z > \sigma_M$ sets a lower bound on the detection volume, $V > \left(\frac{2\ k_B\ T}{\gamma_p\ B_0}\right)^2 \frac{1}{\rho_p} \approx (9\ \mu m)^3$. Here $\gamma_p = 1.41 \times 10^{-26}$ J/T is the proton magnetic moment, $k_B = 1.38 \times 10^{-23}$ J/K is Boltzmann's constant, T = 300 K is the temperature, and $\rho_p = 6.7 \times 10^{28}$ / m$^3$ is the density of protons in water.

We constructed an NV ensemble magnetic sensor, designed to detect NMR signals due to the thermal sample magnetization (Fig. 3a). The sensing volume consisted of the overlap region between a 13 μm thick NV-doped layer (NV density 2 × 10$^{17}$ cm$^{-3}$) at the diamond surface, and the ~10 μm diameter waist of our optical excitation beam, which was totally internally reflected off the diamond surface at an angle of ~45°. Applying a sinusoidal magnetic test signal from a nearby coil antenna at $f$ = 3.742 MHz, we measured an SR sensitivity (using $f_0$ = 3.74066 MHz) of ~50 pT/Hz$^{½}$ [supplementary materials]. By comparison, the expected signal size due to a large sample [$V \gg (9\ \mu m)^3$] of protons in water is ~81 pT [supplementary materials].

To motivate this choice of sensor geometry, we consider a single NV center, located a depth $d$ below the diamond surface, and a sample consisting of a half-space of Larmor-precessing spins (density $\rho_p = 6.7 \times 10^{28}/m^3$) above the surface. Comparing the standard deviation of magnetic field fluctuations at the position of the NV, $\sigma_B$, to the mean magnetic field due to the sample magnetization, $\bar{B}$, we find that the condition for the latter to dominate is $d_{NV} \gtrsim 3\ \mu m$ [supplementary materials]. (Note that both $\sigma_B$ and $\bar{B}$ are obtained by projecting the magnetic field onto the dipole axis of the NV, which has been aligned parallel to the direction of $B_0$.) On the other hand, the effective detection volume for the mean sample magnetization, defined here as the radius $r$ of a hemisphere above the sensor such that spins within the hemisphere contribute exactly half of $\bar{B}$, is given by $V = \frac{2\pi}{3}(\kappa\, d_{NV})^3$, for $\kappa \approx 2.4$ a geometric constant estimated by numerical integration (Fig. 3a inset, and supplementary materials). The present design, with mean NV depth $d = 6.5\ \mu m$, thus represents a tradeoff between (i) suppression of magnetic noise due to near-surface spin fluctuations, and (ii) minimization of the effective sensing volume for the thermal sample magnetization.

Using the NV ensemble sensor, we applied SR spectroscopy to detect NMR signals from glycerol ($C_3H_8O_3$) molecules. The diamond was placed in a cuvette filled with glycerol (volume 0.64 mL) and aligned in the bias field of an electromagnet ($B_0$ = 88 millitesla). At the start of the experiment, a $\pi/2$ pulse was applied to tip the sample protons into the transverse plane of the Bloch sphere; we then measured their precession frequency by probing the NV centers with an SR sequence (parameters $\tau_{SR}$ = 24.06 μs and $n = 4 \times 10^4$). The SR sequence duration was chosen to allow full population relaxation of the sample spins ($T_1 \approx 3 \times 10^{-2}$ s [15]). After $7 \times 10^4$ averages, the nuclear FID was readily observable (Fig. 3b). Near the end of the SR sequence, after the sample spins were fully dephased, we used a coil antenna to drive a calibrated oscillating magnetic field pulse (zero-to-peak amplitude 90 pT, offset frequency $\delta f$ = 1.4 kHz from $f_0$, duration 12 ms). Comparison of integrated peak intensities in the SR power spectrum (Fig. 3b, inset) yielded a signal amplitude of 105 pT (zero-to-peak) for the glycerol FID, approximately consistent with the calculated value of 79 pT for a glycerol proton spin density of $\rho_p = 6.7 \times 10^{28}/m^3$. To exclude the possibility of spurious detection associated with room noise or sensor imperfections, we swept $B_0$ over 0.02 millitesla and repeated the SR protocol at each value. A linear fit to the measured FID line centers gave the correct value for the proton gyromagnetic ratio, $\gamma_p$ = (42.574 ± 0.002) MHz/tesla (Figure 3c).

To assess the spectral resolution limits of NMR detection using this technique, we carried out SR spectroscopy on a sample of pure water. The experimental conditions were identical those of the glycerol measurements, except the full SR sequence duration was extended ($T$ = 2 s) to account for water's longer decoherence and population decay lifetimes ($T_2$, $T_1$ > 2s [14]). The water FID linewidth in the SR power spectrum was 9 ± 1 Hz FWHM (Figure 3d), approximately a factor of ~25 narrower than the best spectral resolution obtained using NV correlation spectroscopy [9]. Nevertheless, the observed lineshape was notably broader than the limiting value ($\Gamma_{FWHM} \approx 0.2$ Hz) associated with the intrinsic bulk decoherence lifetime of the sample.

We therefore investigated a number of effects that could contribute to sample dephasing in the SR spectroscopy of water. Temporal inhomogeneity of the bias field $B_0$ was excluded by active stabilization using the electromagnet current supply, with residual fluctuations < 0.05 μT RMS (equivalent to a proton linewidth $\Gamma$ < 2.5 Hz FWHM) over the course of the experiment [supplementary materials]. Gross spatial gradients in $B_0$ were ruled out by continuous wave electron spin resonance (cw-ESR) measurements [16] at defined positions across the diamond surface, which showed $B_0$ variability < 0.3 µT over an area of ~1 mm$^2$. To check for broadening due to far-detuned proton driving by the NV magnetometry sequence (which should scale approximately as $\Gamma \sim (\Omega_R \cdot \gamma_p / \gamma_{NV})^2 / \Delta \sim 1$ Hz, for $\Omega_R \approx 15$ MHz the NV Rabi frequency, $\Delta \approx 400$ MHz the detuning, and $\gamma_p$ = 42.58 MHz/tesla and $\gamma_{NV}$ = 28.02 GHz/tesla the proton and NV gyromagnetic ratios, respectively), we reduced the NV Rabi frequency by up to a factor of three, but observed no narrowing of the proton resonance [supplementary materials]. Finally, to test whether the proton line was broadened by magnetic field gradients associated with repolarization of the NV electronic spins at each SR readout iteration, (an effect estimated to contribute to broadening at the $\Gamma \sim 1$ Hz level [supplementary materials]), we varied the duty cycle of the magnetometry sub-sequence (relative to the SR cycle period $\tau_{SR}$) from 0.18 (using XY8-2) to 0.53 (using XY8-6) [supplementary materials]. This produced no significant increase in the proton linewidth, indicating that NV back-action was not the dominant dephasing mechanism.

Having thereby ruled out both $B_0$ inhomogeneity and interactions with the NV spins as primary determinants of the observed water FID signal width, we attributed the limited spectral resolution of our measurements to a combination of (i) micron-scale magnetic gradients due to susceptibility differences between sensor components, and (ii) dephasing due to diffusion of sample molecules close to magnetic defects at the diamond surface [17]. We tested this hypothesis by applying $\pi$-pulses to the protons at times $t$ = 40 ms and $t$ = 120 ms after the start of the SR protocol, to refocus proton spin dephasing due to local gradients. This resulted in a proton linewidth of 2.8 Hz FWHM (Figure 3d), approximately consistent with the measured distribution of Gaussian temporal fluctuations in $B_0$ (~2.5 Hz FWHM) recorded by the electromagnet feedback controller over the full duration of the experiment. Gradient-induced broadening is commonly observed in sub-µL volume NMR spectroscopy with microcoils [18,19], and can be mitigated by improved susceptibility matching in the sensor design. Dephasing due to shallow paramagnetic impurities may be reduced by careful diamond surface preparation using a combination of wet-etching and annealing in oxygen [7].

To illustrate the applicability of the NV SR technique to molecular NMR in a picoliter volume, we acquired liquid-state FID spectra of trimethyl phosphate [PO(OCH$_3$)$_3$] and xylene [(CH$_3$)$_2$C$_6$H$_4$]. Trimethyl phosphate (TMP) is a standard reagent known to have large scalar coupling (J[P,H] $\approx$ 11 Hz) between the methyl protons and the central $^{31}$P nuclear spin.[20] Xylene is an aromatic solvent with substantial chemical shifts (~5 ppm of the proton Larmor frequency) due to different electron densities associated with the carbon ring structure and satellite methyl groups. Data were acquired using the same procedure as for glycerol and water FID spectra; the SR protocol parameters were $\tau_{SR}$ = 24 µs and $n = 4 \times 10^4$. The SR NMR spectrum for TMP (Figure 4a) shows two clearly resolved peaks due to the J-coupled nuclei, with splitting $\Delta f_J \approx 13 \pm 1$ Hz. The SR NMR spectrum for xylene (Figure 4b) also shows two peaks, split by $\Delta f_{CSJ} \approx 20 \pm 2$ Hz, consistent with the previously reported [21] value for the chemical shift. The observed peak intensity ratio of ~2.2:1 in the xylene SR NMR power spectrum is as expected for the relative nuclear abundance of 6:4, with the protons in high electron-density methyl groups shifted to lower frequency. These measurements constitute the first demonstration of NV-detected NMR with spectral resolution sufficient to resolve frequency shifts due to molecular structure.

## Discussion:

We have implemented the synchronized readout (SR) protocol [13] using both a single NV at the nanoscale, as well as an NV ensemble sensor that is optimized to detect NMR signals from thermally-polarized samples with volume ~1 picoliter, i.e., $V \approx (10~\mu m)^3$. The NV ensemble SR sensor fills an important technological gap between nanoscale magnetic resonance techniques (e.g., single NV-correlation spectroscopy, $V \approx (5~nm)^3$ [5,6] and magnetic resonance force microscopy, $V \approx (10~nm)^3$ [22]) and detection using inductive micro-coils ($V \gtrsim (100~\mu m)^3$ [22]), in concert with ongoing advances using giant magnetoresistance (GMR) [23] probes. Furthermore, because the NV ensemble SR sensor is not subject to limitations associated with finite NV coherence time or diffusion-limited sample correlation time, it provides spectral resolution nearly two orders of magnitude narrower than previously demonstrated in NV-detected NMR, enabling observation of J-couplings and chemical shifts for the first time using a solid-state spin sensor. Of particular interest in this picoliter sample-volume regime is the possibility of performing NMR spectroscopy of small molecules and proteins [24] at the single-cell level. While some work has been done on inductively-detected intracellular NMR with slurries of bacterial cells [25] and large individual eukaryotic cells such as oocytes [26], NMR spectroscopy of smaller individual cells has not been achieved. By increasing $B_0$ from 88 millitesla to ~1 tesla, both the proton number sensitivity and spectral resolution of an NV ensemble SR sensor can be improved by at least an order of magnitude, which may enable useful single cell NMR.

For nano-NMR applications using single-NV SR, the requirement of weak sample-sensor coupling, combined with imperfect spin state readout of single-NV experiments [27], presents a technical challenge. In particular, weak coupling implies small NV phase accumulation at each magnetometry pulse sequence. However, the inherently statistical nano-NMR signal will be averaged incoherently over many repetitions of the SR protocol, and is added in quadrature with large single-NV readout noise in the SR power spectrum, making the signal difficult to detect. Techniques to improve NV readout fidelity, such as repetitive readouts of a nuclear memory [28] or NV charge-state detection [29], could in principle mitigate the problem. Even so, the challenge of short signal correlation times from nanoscale liquid-state NMR samples due to molecular diffusion is not yet solved, limiting the advantages of a spectrally-selective sensor. Until translational diffusion can be reliably restricted at the nanoscale (e.g., by gel media [**30**] or nanofabricated encapsulation chambers [**31**]) without increased dipolar broadening, SR techniques will likely be of greatest utility in larger (e.g., micron-scale) volumes.

Finally, we consider the possibility of applying SR techniques to a macroscopic NV ensemble sensor for analytical NMR spectroscopy of concentration-limited samples. With a system operating at $B_0 \approx 1$ tesla, the expected proton number sensitivity is ~$10^{13}$ proton spins/Hz$^{\frac{1}{2}}$, which would enable detection of a spin concentration of 1.2 M with SNR $\approx 3$ in 10 minutes of averaging. For applications in chemical analysis, it is desirable to scale the sample region up to $V \approx (1~mm)^3 = 1~\mu L$ to accommodate analyte volumes that can be practically handled in the lab. By using a 1 mm$^3$ diamond (with similar NV density and coherence times to the present sensor), increasing the optical excitation intensity to 1 W, and employing a light-trapping diamond waveguide geometry [32], it should be feasible to obtain a magnetic field sensitivity of ~2 pT/Hz$^{\frac{1}{2}}$. This would enable detection of ~50 mM proton concentrations with SNR $\approx 3$ in 10 minutes of averaging, approaching the concentration sensitivity demonstrated with state of the art microcoils operating at $B_0 = 7 - 12$ tesla [33]. The NV detector sensitivity might be further improved with advances in diamond engineering allowing preferential NV orientation [34] and/or improved N-NV conversion efficiency [35]. Operation of an NV sensor at ten times lower $B_0$ compared to microcoils could be advantageous, potentially mitigating the challenges associated with susceptibility mismatches encountered in many microcoil experiments. Furthermore, the diamond platform should be amenable to parallel operation, using an array of chips with independent (cross-talk free) optical readouts for each. This opens the possibility of parallelized, high-throughput analytical NMR spectroscopy for concentration-limited samples.

**Note**: During preparation of this manuscript, we became aware of two other studies demonstrating the synchronized readout technique for a single NV using a coil-generated magnetic signal [36,37].

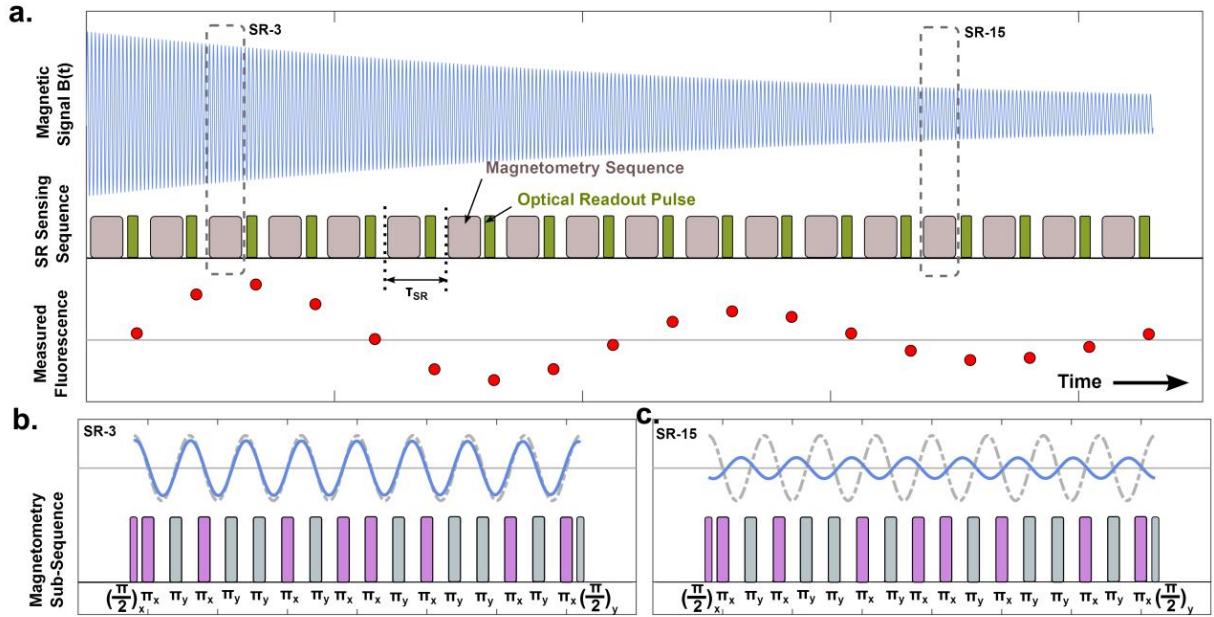

**Fig. 1: Principle of Synchronized Readout (SR) protocol: (a)** Numerical simulation of SR detection of a free induction decay (FID) signal, $B(t)$ (blue), which oscillates at frequency $f$ and has finite decay lifetime $\tau$. The SR sequence consists of interspersed blocks of identical NV magnetometry sub-sequences (grey boxes) and optical NV spin state readouts (green boxes). Using magnetometry sub-sequences with maximum response at frequency $f_0$, the duration $\tau_{SR}$ of each SR iteration is chosen to be $\tau_{SR} = k/f_0$, for integer $k$. The NV fluorescence time series over successive SR readouts oscillates at frequency $\delta f = f - f_0$, because the FID signal phase advances incrementally relative to the magnetometry sub-sequence. **(b)** Detail of calculated magnetic signal and magnetometry subsequence at the third SR iteration (denoted SR-3). The signal (blue line) is nearly in phase with a sinusoid at $f_0$ (grey dashed line). The magnetometry subsequence (here implemented as an XY8-2 dynamical decoupling sequence) consists of a series of π-pulses timed to coincide with the zero-crossings of the sinusoid at $f_0$, resulting in a detected fluoresce maximum because the FID is in phase. **(c)** Detail of magnetic signal and magnetometry subsequence at SR-15. The signal (blue line) has advanced and is now ~180° out of phase with the sinusoid at the central frequency (grey dashed line). This gives rise to a detected fluorescence minimum at SR-15.

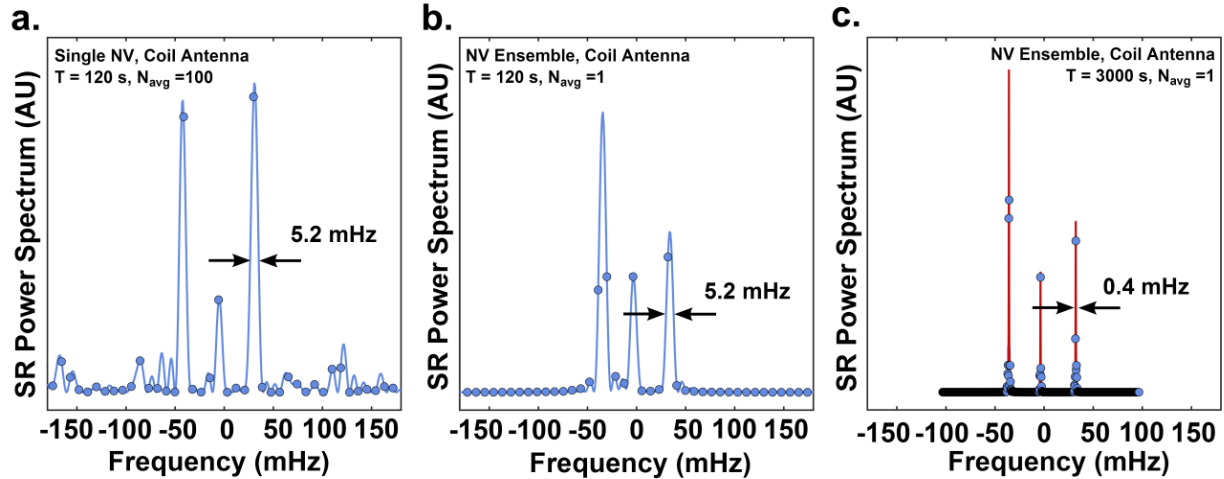

**Fig 2: Measured SR spectral resolution using signals from a coil antenna. (a)** Power spectrum of SR signal obtained with a single-NV magnetic sensor in a confocal microscope. The SR protocol used CPMG-32 magnetometry sub-sequences, with SR iteration time $\tau_{SR}$ = 75 µs, and the total experiment duration was $T = n\,\tau_{SR}$ = 112.5 s, for $n = 1.5 \times 10^6$ iterations. Data shown are the average of 100 experiments. The observed spectral width was 5.2 mHz (FWHM). Independent, spectrally narrow signal sources were used to drive each of the three detected frequencies. **(b)** Power spectrum of SR signal obtained with an NV ensemble magnetic sensor. The SR protocol used XY8-4 magnetometry sub-sequences, with SR iteration time $\tau_{SR}$ = 75 µs, and the total experiment duration was $T = n\,\tau_{SR}$ = 112.5 s, for $n = 1.5 \times 10^6$ iterations. The spectrum shown is for a single average. The observed spectral width was again 5.2 mHz (FWHM). **(c)** Power spectrum of SR signal obtained with an NV ensemble magnetic sensor. The SR protocol used XY8-4 magnetometry sub-sequences, with SR iteration time $\tau_{SR}$ = 1.2 ms, and the total experiment duration was $T = n\,\tau_{SR}$ = 3000 s, for $n = 2.5 \times 10^6$ iterations. The observed spectral width was 0.4 mHz (FWHM). The measured linewidths for all three signal peaks were consistent to within ~10%, suggesting that the spectral resolution was limited by the stability of the timing source used to control the SR protocol, rather than individual signal sources.

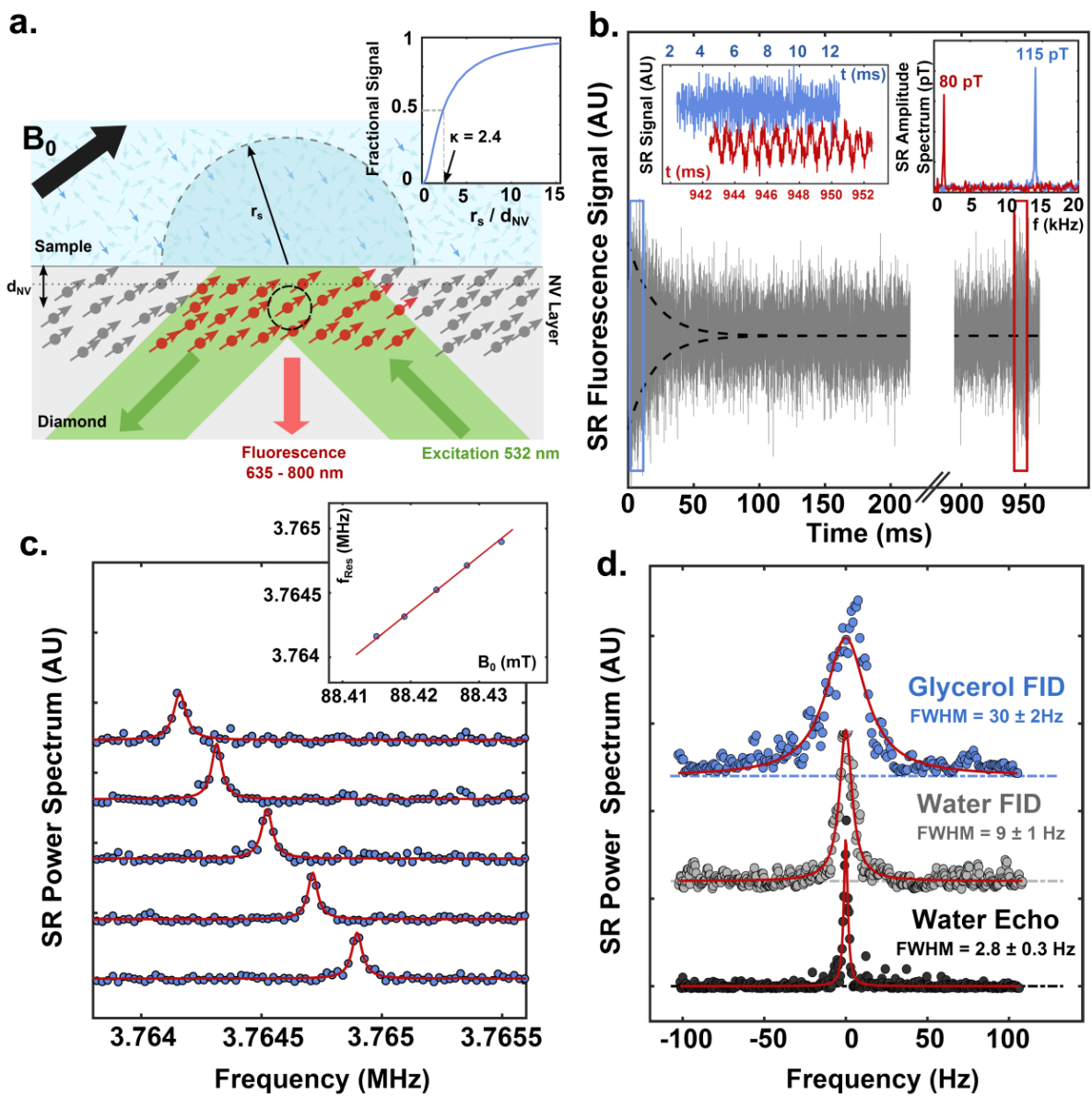

**Figure 3: NV ensemble sensor for SR-detection of proton NMR. (a)** NV ensemble sensor geometry. The sensor consists of a ~13 μm deep layer of NV centers at the surface of a diamond chip, probed by a green excitation laser with beam diameter ~10 μm. The excitation beam is totally internally reflected in the diamond, and the NV spin-dependent fluorescence is collected from the back of the diamond and detected with a photodiode. The sensor is designed to detect NMR from a small, thermally-polarized fraction (~$3 \times 10^{-7}$, dark blue arrows) of the total sample spin population (light blue arrows). The thermally-polarized spins are driven with a resonant coil and precess around the static bias field $B_0$. The NV centers at depths $d_{NV} > 3$ μm from the diamond surface (dotted horizontal line) are primarily sensitive to the thermal spin polarization; shallower NV centers have signals dominated by statistical spin fluctuations. For an NV in the sensing layer at depth $d_{NV}$ (e.g., NV in dashed circle), approximately half of the signal from thermally-polarized spins is due to those in a hemispherical sample volume $r_s < 2.4$ $d_{NV}$ (for $r_s$ indicated by grey semicircle) with the remainder of the spins in a semi-infinite volume contributing the other half of the signal. **(b)** Detection of NMR from protons in glycerol. Grey time trace shows the SR time-series signal produced by FID of glycerol spins above the diamond. Dashed line shows exponential decay envelope obtained by fitting a Lorentzian lineshape to the Fourier transform of the time series data. A calibrated (80 pT zero to peak amplitude) magnetic field from a coil antenna is turned on at around t = 950 ms. Comparison of the FID and antenna signals in the time and/or frequency domains (insets) yields a proton signal amplitude of 115 pT. **(c)** Power spectra of proton resonance frequencies obtained from glycerol FID data (blue circles) for various values of $B_0$, fit to Lorentzian lineshape (solid red lines). A linear fit of resonance frequency vs. $B_0$ (inset) gives the correct proton gyromagnetic ratio, $\gamma_p$ = (42.574 ± 0.002) MHz/T. **(d)** Resolved power spectra obtained by SR-FID from protons in glycerol (blue circles) and pure water (grey circles), as well as by SR-spin-echo in pure water (black circles). The spectral resolution obtained with SR-FID of glycerol was 30 ± 2 Hz (FWHM), as determined by least-squares fitting to a Lorentzian line shape (red line). The spectral resolution obtained from pure water was 9 ± 1 Hz (FWHM) with SR-FID, and 2.8 ± 0.3 Hz (FWHM) with SR-spin-echo.

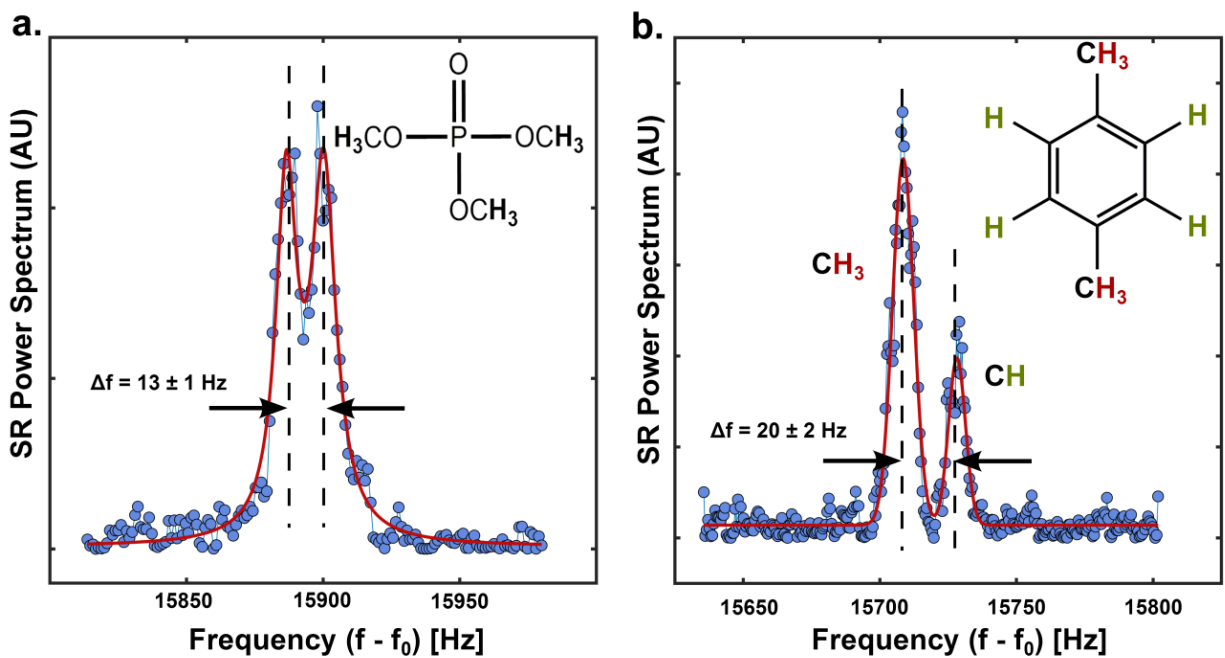

**Figure 4: SR-detected molecular NMR spectra. (a)** SR-FID spectrum of trimethyl phosphate (blue circles), acquired using the SR protocol with XY8-6 magnetometry subsequences, and $\tau_{SR}$ = 24 μs and $n = 4 \times 10^4$. Fitting to a sum of two Lorentzian lineshapes (solid red line) indicates a splitting $\Delta f$ = 13 ± 1 Hz due to J-coupling between the central $^{31}$P nucleus and the methyl protons. **(b)** SR-FID spectrum of xylene (blue circles), acquired using SR with XY8-6 subsequences, and $\tau_{SR}$ = 24 μs and $n = 4 \times 10^4$. The relative peak heights obtained from fits to a sum of Gaussian lineshapes (solid red lines) are due to the relative abundances of CH and CH$_3$ protons in the molecule, and the splitting $\Delta f$ = 20 ± 2 Hz is the result of chemical shifts associated with the two proton positions.

## Methods:

**Diamond Samples:** The diamond used to construct the NV NMR sensor was a 2 mm × 2 mm × 0.5 mm high-purity chemical vapor deposition (CVD) diamond chip, with 99.999% 12C isotopic purity, and bulk nitrogen concentration [$^{14}$N] < 8.5 × 10$^{14}$ cm$^{-3}$ (Element Six). Modification of the CVD gas mix during the final stage of growth yielded a 13 μm thick nitrogen-enriched top layer ([$^{14}$N] ≈ 4.8 × 10$^{18}$ cm$^{-3}$, measured by secondary ion mass spectrometry). The diamond was electron-irradiated (1.3 × 10$^{14}$ cm$^{-2}$ s$^{-1}$ flux) for 5 hours, and annealed in vacuum (800° C) for 12 hours, yielding an NV concentration [NV] ≈ 3 × 10$^{17}$ cm$^{-3}$. The diamond was cut such that the top face was perpendicular to the [100] crystal axis, and the lateral faces were perpendicular to [110]. All four edges of the top face were then polished through at 45° (Delaware Diamond Knives), resulting in a truncated square pyramid, with top face area 1 mm × 1 mm. The ensemble $T_2^*$ dephasing time for NV centers in this diamond, measured using Ramsey spectroscopy, was $T_2^*$ ≈ 750 ns. The ensemble $T_2$ decoherence time, measured using a Hahn-echo sequence, was 6.5 μs.

The diamond used for NV ensemble sensing of antenna signals (Fig 1b -1c) was identical to the NV NMR sensor chip, but without angle-polished edges. The diamond used in single NV experiments was a 4 mm x 4 mm x 0.5 mm high-purity CVD diamond chip, with 99.99% $^{12}$C isotopic purity near the surface, which contained preferentially oriented NV centers with nitrogen concentration [$^{14}$N] ≈ 1 × 10$^{15}$ cm$^{-3}$ and NV concentration [NV] ≈ 3 × 10$^{12}$ cm$^{-3}$. The approximate coherence times for the single NV center used in our experiments were $T_1$ ≈ 1ms, $T_2$ ≈ 500 μs, and $T_2^*$ ≈ 50 μs.

**Single-NV Sensor:** The single-NV sensor was based on a low NV density diamond chip, as described above. The antenna-generated magnetic signals were measured using a home-built scanning laser microscopy system. Confocal scanning of the diamond chip was done by a three-axis motorized stage (Micos GmbH). A 400 mW, 532 nm diode-pumped solid state laser (Changchun New Industries) was used as an excitation light, and an acousto-optic modulator (Isomet Corporation) operated at 80 MHz was used to time-gate the laser. An oil-immersion objective (100x, 1.3 NA, Nikon CFI Plan Fluor) focused the green laser pulses onto a single NV center. Red fluorescence from the NV centers was collected by a silicon avalanche photodetector (Perkin Elmer SPCM-ARQH-12) through a 75μm sized pinhole. The NV spin initialization and readout pulses were 3 μs and 0.5 μs, respectively. For the single-NV magnetometry, microwave pulses were applied to NV centers through nanofabricated 20 μm gapped waveguide, and the pulse sequence was generated using a GHz range signal source (Agilent E4428C), which was I/Q modulated by an AWG (Tektronix 5014c) for the microwave phase control.

**NV Ensemble Sensor:** The NV ensemble sensor was based on a diamond chip with 13 μm NV-enriched surface layer, as described above. For detecting antenna-generated magnetic signals, the diamond chip was rectangular. For NMR sensing, an angle-polished chip was employed instead, allowing total internal reflection of the laser beam (Fig. 3a) to prevent direct illumination of the NMR sample. Excitation light was provided by a diode-pumped solid state laser at 532 nm (Coherent Verdi G7), directed through an acousto-optic modulator (AOM) (IntraAction ASM802B47) to produce 5 μs pulses. The first ~1 μs of each pulse was used to optically read the spin state of the NV ensemble, while the remainder of the pulse repolarized the NVs. The AOM was driven by a digitally synthesized 80 MHz sinusoid (Tektronix AWG 7122C), amplified to 33 dBm (Minicircuits ZHL-03-5WF), and the total laser power at the sensor volume was 150 mW. The laser was focused to a 20 μm diameter waist near the position of the NV sensor layer, resulting in an optical intensity 48 kW/cm$^2$ (comparable to the NV saturation intensity of 100 kW/cm$^2$).

For detection of antenna magnetic signals, the diamond was mounted on a glass slide; for NMR detection, it was glued (Epoxy Technology Inc., EPO-TEK 301) to a 3 mm glass prism (Thorlabs PS905) and placed inside a sample cuvette (FireflySci Type4 Microfluorescence Cuvette). In both cases, the diamond was carefully rotated such that a [111] diamond crystal axis was aligned to the static magnetic field $B_0$. NV centers aligned along this axis were used for sensing, while those along the other three [111] directions were far off-resonance and contributed only to the background fluorescence. The alignment was carried out by overlapping the pulsed electron spin resonance (ESR) frequencies of the 3 non-aligned axes. The static magnetic field strength was $B_0$ = 88 millitesla, such that the resonance frequency of the $|m_s = 0\rangle \rightarrow |m_s = -1\rangle$ spin transition for the aligned NV centers was $f_{Larmor}$ = 400 MHz. (The $|m_s = 0\rangle \rightarrow |m_s = +1\rangle$ resonance frequency was 5340 MHz.)

The NV magnetometry pulse sequences for magnetic resonance detection were carried out on the $|m_s = 0\rangle \rightarrow |m_s = -1\rangle$ transition. Microwaves were delivered using a straight length of wire (0.25 mm diameter) above the diamond, approximately 0.4 mm away from the NV sensing volume. Both the 400 MHz carrier frequency and the pulse modulation were synthesized digitally (Tektronix AWG 7122C); pulses were then amplified to 40 dBm (Minicircuits ZHL-100W-52-S+) and coupled into the wire, yielding NV Rabi frequency $\Omega$ = 8.3 MHz. An XY8-6 dynamical decoupling sequence was used to selectively detect magnetic resonance signals around 3.755 MHz, the which is the proton Larmor frequency at $B_0$ = 88 millitesla. The phase of the final π/2 pulse of the sequence was optimized to give fluorescence corresponding to a mixed state of the NV (i.e. equal to the mean fluorescence over one Rabi oscillation), to make the fluorescence signal linearly sensitive to small magnetic field amplitudes. For an ideal two-level quantum system, this condition would correspond to a 90° phase shift between the initial and final π/2 pulses; in practice, small drive detunings associated with $^{14}$NV hyperfine structure required manual optimization of the phase. To reject laser intensity noise and microwave power fluctuations, the phase of the final π/2 pulse of every second SR magnetometry subsequence was shifted by 180° relative to the nominal value, and successive pairs of readouts were amplitude-subtracted. Thus, one SR time-series data point was recorded for every two magnetometry subsequences.

Spin state-dependent fluorescence from the NV centers was collected with a quartz light guide (Edmund Optics 5mm Aperture, 120 mm L, Low NA Hexagonal Light Pipe) and delivered to a balanced photodiode module (Thorlabs PDB210A). To eliminate scatter from the excitation laser, an interference filter (Semrock BLP01-647R) was placed between the light guide and detector. A small fraction of the excitation beam was split off upstream of the diamond chip and directed onto the second channel of the balanced diode module. A glass slide mounted on a motorized stage (Thorlabs PRM1Z8) in the second path allowed automated re-balancing between averages during long SR signal acquisitions. When the NV centers were fully polarized in $|m_s = 0\rangle$, the light-induced fluorescence signal produced a single-channel (unbalanced) photocurrent of 30 μA. Immediately after applying a microwave π pulse, the single-channel photocurrent was 28 μA, indicating a maximum fluorescence contrast of ∼7%. The difference signal of the photodiode module (with onboard transimpedance gain 1.75 × 10$^5$ V/A) was further amplified by 3 dB and low-pass filtered at 1 MHz using a low-noise pre-amplifier unit (Stanford Research SR-560), then recorded with a digital to analog converter (DAQ) (National Instruments NI-USB 6281). The DAQ bandwidth was 750 kHz, and the digitization was on-demand, triggered by a TTL pulse from the AWG used to control the experiment. The delay between the rising edges of the AOM gate pulse and the DAQ trigger was 1.9 µs, optimized for maximum spin state-dependent fluorescence contrast.

**SR Protocol Synchronization and Data Analysis:** The SR cycle period $\tau_{SR}$ = 24.06 µs, the reciprocal central SR detection frequency, $f_0^{-1}$ = 1 / (3.74065 MHz) = 267.3 ns, and the reciprocal NV drive frequency, $f_{Larmor}^{-1}$ = 1/(400 MHz) = 2.5 ns, were all chosen to be exact integer multiples of the clock period of the timing generator (Tektronix AWG 7122C), $\tau_{Clock}$ = (1/12 GHz) = 0.083 ns. The ultimate frequency resolution of the experiment was therefore determined by the stability of this clock. The NV magnetometry pulse sequence (XY8-6 in all experiments, unless otherwise specified in the main text) was saved in the memory of the AWG and its output was gated by a TTL signal from a programmable pulse generator (Spincore PulseBlasterESR-PRO 500 MHz). The PulseBlaster gate duration was used to specify $n$, the number of SR iterations per experiment. For detecting the NMR signals, the pulse blaster also generated the TTL pulse for gating the proton driving MW pulses. Each readout of the SR protocol was saved in a numerical array, giving a time series of length $n$. Individual time series were averaged (in the time domain) to improve SNR. The first 20 SR time series data points, which coincided with the proton pulse $\pi/2$ pulse plus approximately 50 times the coil ringdown time, were discarded. The averaged time series data were then mean-subtracted before taking the Fourier transformation and fitting using Matlab. Each spectrum was fit to both Lorentzian and Gaussian lineshapes, and the model with smaller residuals (always Lorentzian, except in the case of Figure 4b) was selected for display. Unless otherwise specified, all spectra shown in the figures are power spectra, calculated as the absolute value of the Fourier-transformed time series data. When uncertainties are quoted for spectral linewidth or splitting parameters, these uncertainties were estimated by repeating the full experiment and fitting procedure several times, then calculating the standard deviation over the ensemble of fitted parameters.

**Electromagnet**: The bias magnetic field $B_0$ was produced by an air-cooled electromagnet (Newport Instruments Type A). The pole pieces were cylindrical, 10 cm in diameter, with adjustable gap set to 3 cm. The main coils (each 1900 turns of copper strip, with room-temperature resistance R = 4.5 Ω) were driven (Hewlett Packard HP 6274) with a continuous current of ~650 mA to produce a nominal field $B_0 \approx$ 88 millitesla. A secondary coil pair (diameter 10 cm, gap 7 cm, 15 turns each) were manually wound around the poles to allow precise field stabilization without the need for very small adjustments to main current supply. The secondary coils were driven by a voltage-controlled current supply (Thorlabs LDC205C), controlled by the analog output channel of a DAQ (National Instruments PCI 6036E). The field strength was monitored using continuous wave electron spin resonance (cwESR) measurements on a secondary diamond chip, spatially separated from the main sensor by ~1 cm [supplementary materials]. The cwESR microwave frequency modulation was synchronized to the main SR experiment using the same AWG (Tektronix AWG 7122C), to ensure that any cross-talk between the detectors was coherent over averages of the SR protocol and could be removed during data analysis. (This precaution proved unnecessary in the final experiments, as the cwESR drive power was too weak to produce a measureable effect on the SR sensor.) The excitation laser, light collection optics, and microwave drive for the secondary experiment were all independent from those of the main SR magnetic resonance sensor. This enabled feedback control over magnetic field fluctuations (primarily due to current noise in the main coils) with bandwidth ~12.5 Hz, resulting in short-term (~30 minutes) field stability $\sigma_B \approx 30$ nT RMS. To correct slow drifts between the main magnetic resonance sensor and the secondary field-stabilization sensor, we periodically (every 5 minutes) paused the SR protocol between averages and performed pulsed ESR measurements on the primary diamond. Any measured magnetic field drifts were used to correct the setpoint of the fast feedback loop, ensuring long-term (~50 hours) stability $\sigma_B \approx 50$ nT RMS. All cwESR measurements were carried out using both the $|m_s = 0\rangle \rightarrow |m_s = -1\rangle$ and the $|m_s = 0\rangle \rightarrow |m_s = +1\rangle$ transitions of the aligned NV centers, to distinguish resonance shifts due to changes in temperature [38] and magnetic field. For fast feedback measurements on the secondary sensor, we monitored only 4 discrete ODMR frequencies to maximize bandwidth. This system was potentially susceptible to second-order feedback errors associated with simultaneous changes in $B_0$ and temperature. We therefore thermally anchored the secondary sensor to a piece of black-anodized aluminum and actively stabilized its temperature using absorption from a separate DPSS laser (Thorlabs DJ532-40). Temperature control was not required for slow feedback on the main magnetic resonance sensor, where we acquired a full ODMR spectrum (58 frequency points) to fully account for all drifts in magnetic field, temperature, and optical contrast.

**NMR Drive Coils:** Radio frequency (RF) pulses for driving sample protons (e.g., with a $\pi/2$-pulse at the start of an SR-FID experiment) were produced by a pair of solenoid coils wound around the sample cuvette. This geometry, with 1.1 cm coil diameter and 1.2 cm center-to-center spacing, provided a combination of strong drive fields and convenient optical access to the NV ensemble sensor. The coils were 24 turns each, connected in series and coupled to the current source (Rigol DG 1032) with a standard network of variable matching and coupling capacitors [39]. After tuning, the resonance frequency was 3.75 MHz, and the coil Q was 140. Driving the coils on resonance, we obtained a maximum nuclear Rabi frequency $\Omega$ = 2.8 kHz.

**NMR Samples:** Deionized water was obtained from Ricca Chemical Company (part number 9150-5). p-Xylene, Glycerol and Trimethylphosphate were purchased from Sigma Aldrich and used without dilution or modifications (Sigma-Aldrich catalog numbers 296333, G9012 and 241024 respectively). The glycerol sample may have contained some atmospherically-absorbed water (< 20% by volume).

## Acknowledgements:

This material is based upon work supported by, or in part by, the U. S. Army Research Laboratory and the U. S. Army Research Office under contract/grant number W911NF1510548. D.B.B. was partially supported by the German Research Foundation (BU 3257/1-1). We thank Roger Fu for assisting with acquisition of the electromagnet used to create the applied bias field; and Matthew Rosen for guidance on NMR techniques.

# References:

[1] H. Guenther, *NMR Spectroscopy Basic Principles, Concepts and Applications in Chemistry* (Wily-VCH, Weinheim, Germany, ed. 3, 2013).

[2] M. E. Lacey, R. Subramanian, D. L. Olson, A. G. Webb, J. V. Sweedler, High-Resolution NMR Spectroscopy of Sample Volumes from 1 nL to 10 µL. *Chem. Rev*. **99**, 3133 – 3152 (1999).

[3] M. P. Augustine, D. M. TonThat, J. Clarke, SQUID detected NMR and NQR, *Solid State Nucl. Magn. Reson.* **11**, 139 – 156 (1998).

[4] I. M. Savukov, M. V. Romaliz, NMR Detection with an Atomic Magnetometer, *Phys. Rev. Lett.* **94**, 123001 (2005).

[5] H. J. Mamin, M. Kim, M. H. Sherwood, C. T. Rettner, K. Ohno, D.D. Awschalom, D. Rugar, Nanoscale Nuclear Magnetic Resonance with a Nitrogen-Vacancy Spin Sensor, *Science* **339**, 557 – 560 (2013).

[6] T. Staudacher, F. Shi, S. Pezzagna, J. Meijer, J. Du, C. A. Meriles, F. Reinhard, J. Wrachtrup, Nuclear Magnetic Resonance Spectroscopy on a (5-Nanometer)$^3$ Sample Volume, *Science* **339**, 561 – 563 (2013).

[7] I. Lovchinsky, A. O. Sushkov, E. Urbach, N. P. de Leon, S. Choi, K. De Greve, R. Evans, R. Gertner, E. Bersin, C. Mueller, L. McGuinness, F. Jelezko, R. L. Walsworth, H. Park, M. D. Lukin, Nuclear magnetic resonance detection and spectroscopy of single proteins using quantum logic, *Science* **351**, 836 – 841 (2016).

[8] T. Staudacher, N. Raatz, S. Pezzagna, J. Meijer, F. Reinhard, C. A. Meriles, J. Wrachtrup, Probing molecular dynamics at the nanoscale via an individual paramagnetic centre, *Nat. Commun.* **6**, 8257 (2015).

[9] S. Zaiser, T. Rendler, I. Jakobi, T. Wolf, S.-Y. Lee, S. Wagner, V. Bergholm, T. Schulte-Herbrueggen, P. Neumann, J. Wrachtrup, Enhancing quantum sensing sensitivity by a quantum memory, *Nat. Commun.* **7** 12279 (2016).

[10] T. Rosskopf, J. Zopes, J. M. Boss, C. L. Degen, A quantum spectrum analyzer enhanced by a nuclear spin memory. Preprint at https://arxiv.org/abs/1610.03253 (2016).

[11] B. E. Herzog, D. Cadeddu, F. Xue, P. Peddibhotia, M. Poggio, Boundary between the thermal and statistical polarization regimes in a nuclear spin ensemble, *Appl. Phys. Lett.* **105**, 043112 (2014).

[12] C. A. Meriles, L. Jiang, G. Goldstein, J. S. Hodges, J. Maze, M. D. Lukin, P. Cappellaro, Imaging mesoscopic nuclear spin noise with a diamond magnetometer, *J. Chem. Phys.* **133**, 124105 (2010).

[13] R. L. Walsworth, D. R. Glenn, D. Bucher, Synchronized-Readout for narrowband detection of time-varying electromagnetic fields using solid state spins, PCT/US17/34256, Application Number: 62/341,497, filed May 25, 2016.

[14] D. Farfurnik, A. Jarmola, L. M. Pham, Z. H. Wang, V. V. Dobrovitski, R. L. Walsworth, D. Budker, N. Bar-Gill, Optimizing a dynamical decoupling protocol for solid-state electronic spin ensembles in diamond, *Phys. Rev. B* **92**, 060301(R) (2015).

[15] M. Bloembergen, E. M. Purcell, R. V. Pound, Relaxation Effects in Nuclear Magnetic Resonance Absorption, *Phys. Rev.* **73**, 679 (1948).

[16] A. Dreau, M. Lesik, L. Rondin, P. Spinicelli, O. Arcizet, J.-F. Roch, V. Jacques, Avoiding power broadening in optically detected magnetic resonance of single NV defects for enhanced dc magnetic field sensitivity, *Phys. Rev. B* **84**, 195204 (2011).

[17] M. S. Grinolds, M. Warner, K. De Greve, Y. Dovzhenko, L. Thiel, R. L. Walsworth, S. Hong, P. Maletinsky, A. Yacoby, Subnanometre resolution in three-dimensional magnetic resonance imaging of individual dark spins, *Nat. Nanotechnol.* **9**, 279 – 284 (2014).

[18] D. L. Olson, T. L. Peck, A. G. Webb, R. L. Magin, J. V. Sweedler, High-Resolution Microcoil 1H-NMR for Mass-Limited, Nanoliter-Volume Samples, *Science* **270**, 1967 – 1970 (1995).

[19] R. M. Fratila, A. H. Velders, Small-Volume Nuclear Magnetic Resonance Spectroscopy, *Annu. Rev. Anal. Chem.* **4**, 227 – 249 (2011).

[20] S.-H. Liao, M.-J. Chen, H.-C. Yang, S.-Y. Lee, H.-H. Chen, H.-E. Horng, S.-Y. Yang, A study of J-coupling spectroscopy using the Earth's field nuclear magnetic resonance inside a laboratory, *Rev. Sci. Instrum.* **81**, 104104 (2010).

[21] N. R. Babij, E. O. McCusker, G. T. Whiteker, B. Canturk, N. Choy, L. C. Creemer, C. V. De Amicis, N. M. Hewlett, P. L. Johnson, J. A. Knobelsdorf, F. Li, B. A. Lorsbach, B. M. Nugent, S. J. Ryan, M. R. Smith, Q. Yang, NMR Chemical Shifts of Trace Impurities: Industrially Preferred Solvents Used in Process and Green Chemistry, *Org. Process Res. Dev.* **20**, 661 – 667 (2016).

[22] C. L. Degen, M. Poggio, H. J. Mamin, C. T. Rettner, D. Rugar, Nanoscale magnetic resonance imaging, *Proc. Natl. Acad. Sci USA* **106**, 1313 – 1317 (2009).

[23] P. A. Guitard, R. Ayde, G. Jasmin-Lebras, L. Caruso, M. Pannetier-Lecoeur, C. Fermon, Local nuclear magnetic resonance spectroscopy with giant magnetic resistance-based sensors, *Appl. Phys. Lett.* **108**, 212405 (2016).

[24] C. Masoulis, X. Xu, D. A. Reiter, C. P. Neu, Single Cell Spectroscopy: Noninvasive Measures of Small-Scale Structure and Function, *Methods* **64**, 119 – 128 (2013).

[25] Z. Serber, L. Corsini, F Durst, V. Doetsch, In-Cell NMR Spectroscopy, *Methods Enzymol.* **394**, 17 – 41 (2005).

[26] M. Grisi, F. Vincent, B. Volpe, R. Guidetti, N. Harris, A. Beck, G. Boero, NMR Spectroscopy of single sub-nL ova with inductive ultra-compact single-chip probes

[27] M. W. Doherty, N. B. Manson, P. Delaney, F. Jelezko, J. Wrachtrup, L. C. L. Hollenberg, The nitrogen-vacancy colour center in diamond, *Phys. Rep.* **528**, 1 – 46 (2013).

[28] L. Jiang, J. S. Hodges, J. R. Maze, P. Maurer, J. M. Taylor, D. G. Cory, P. R. Hemmer, R. L. Walsworth, A. Yacoby, A. S. Zibrov, M. D. Lukin, Repetitive Readout of a Single Electronic Spin via Quantum Logic with Nuclear Spin Ancillae, *Science* **326**, 267 – 272 (2009).

[29] B. J. Shields, Q. P. Unterreithmeier, N. P. de Leon, H. Park, M. D. Lukin, Efficient Readout of a Single Spin State in Diamond via Spin-to-Charge Conversion, *Phys. Rev. Lett.* **113**, 136402 (2015).

[30] R. M. Dickson, D. J. Norris, Y.-L. Tzeng, W. E. Moerner, Three-Dimensional Imaging of Single Molecules Solvated in Pores of Poly(acrylamide) Gels, *Science* **274**, 966-968 (1996).

[31] M. J. Shon, A. E. Cohen, Mass Action at the Single-Molecule Level, *J. Am. Chem Soc.* **134**, 14618-14623 (2012).

[32] H. Clevenson, M. E. Trusheim, C. Teale, T. Schroeder, D.Braje, D. Englund, Broadband magnetometry and temperature sensing with a light-trapping diamond waveguide, Nat. Phys. 11, 393 – 397 (2015).

[33] V. Badilita, R. Ch. Meier, N. Spengler, U. Wallrabe, M. Utz, J. G. Korvink, Microscale nuclear magnetic resonance: a tool for soft matter research, Soft Matter 8, 10583 – 10597 (2012).

[34] H. Ozawa, K. Tahara, H. Ishiwata, M. Hatano, T. Iwasaki, Formation of perfectly aligned nitrogen-vacancy ensembles in chemical-vapor-deposition-grown diamond (111), *Appl. Phys. Express* **10**, 045501 (2017).

[35] G. Kucsko, S. Choi, P. C. Maurer, H. Sumiya, S. Onoda, J. Isoya, F. Jelezko, E. Demler, N. Y. Yao, M. D. Lukin, Critical thermalization of a disordered dipolar spin system in diamond. Preprint at https://arxiv.org/abs/1609.08216 (2016).

[36] S. Schmitt, T. Gefen, F. M. Stürner, T. Unden, G. Wolff, C. Müller, J. Scheuer, B. Naydenov, M. Markham, S. Pezzagna, J. Meijer, I. Schwarz, M. Plenio, A. Retzker, L. McGuinness, F. Jelezko, Submillihertz magnetic spectroscopy performed with a nanoscale quantum sensor, *Science* **351**, 832 – 837 (2017).

[37] J. M. Boss, K. S. Cujia, J. Zopes, C. Degen, Quantum sensing with arbitrary frequency resolution, *Science* **356**, 837 – 840 (2017).

[38] V. M. Acosta, E. Bauch, M. P. Ledbetter, A. Waxman, L. S. Bouchard, D. Budker, Temperature Dependence of the Nitrogen-Vacancy Magnetic Resonance in Diamond, *Phys. Rev. Lett.* **104**, 070801 (2010).

[39] D. D. Wheeler, M. S. Conradi, Practical Exercises for Learning to Construct NMR/MRI Probe Circuits, *Concepts Magn. Reson. Part A* **40**, 1 – 13 (2012).

## Supplementary Figure 1: Diamond Ensemble Magnetometer SR Sensitivity

In order to determine the magnetic field sensitivity of the NV ensemble magnetometer, we calibrated the amplitude $b_{ac}$ of a test field applied using a nearby coil antenna:

$$b(t) = b_{ac}\sin(2\pi f_{coil})$$

We varied $b_{ac}$ by adjusting the current supplied to the coil antenna at $f_{coil}$ = 3.742 MHz, and monitored the detected SR signal (Figure S1a). The voltage shown here is the control voltage of our signal source, which is linearly proportional to the current fed to the antenna. The voltage was increased while recording SR time series data, until the peak of the SR signal reached saturation and folded back. At that point the NV ensemble had accumulated a phase of 90 degrees during a single NV magnetometry sequence.

We can calculate the $b_{ac}$ field strength which is necessary for acquiring $2\pi$ phase accumulation by integrating the oscillating field during our XY8-6 sequence which gives:

$$b_{ac} = \frac{2\hbar\pi^2 f}{g\mu_B N}$$

where g is the g-factor, $\mu_B$ the Bohr magneton and N the number of pi pulses. For our XY8-6, $2\pi$ phase accumulation on the NV equals to a $b_{ac}$-field of 8.75 µT. Fitting the oscillation in figure S1(b) to a sinusoid gives a period of 1.6 Vpp voltage which results in 5.5 µT/Vpp. Using 2 mVpp generates a field of 11 nT which is detected in 1 second, as shown in figure S1c. The noise floor is about 50 pT which results in a sensitivity of 50 picotesla/Hz$^{½}$.

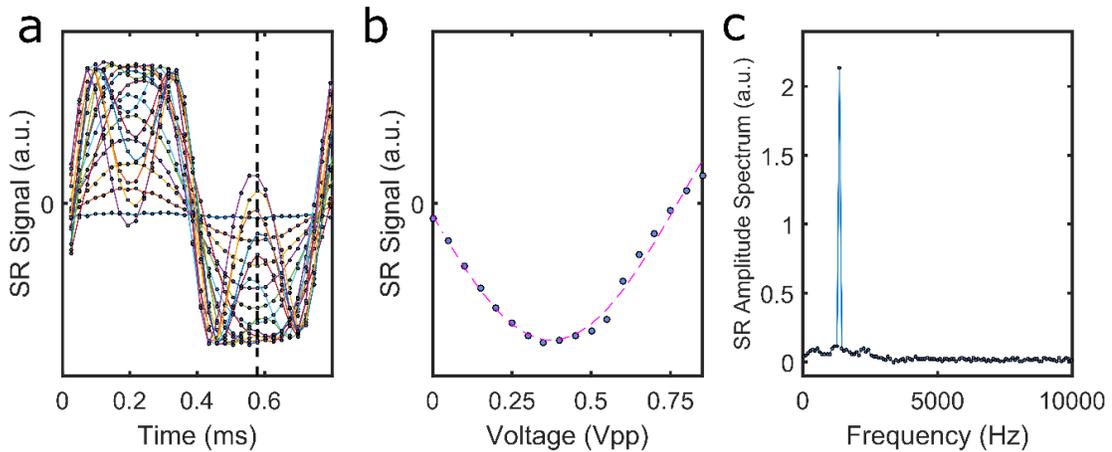

**Figure S1: Estimated sensitivity of the NV ensemble sensor. (a)** Synchronized readout signal for different external magnetic test signal strengths from a nearby coil antenna at $f$ = 3.742 MHz. The control voltage of the coil AC current source is changed from 0 – 0.75 V$_{p-p}$. Each colored line corresponds to a different value of the driving voltage. **(b)** SR signal as a function of coil drive voltage at constant time, obtained by cutting the data of **(a)** along the dashed line. **(c)** 11 nT test signal recorded in 1 second.

## Supplementary Note 2: Calculated NMR Magnetic Field Amplitudes

We would like to estimate the magnetic field produced in the NV ensemble sensor by a sample of proton spins. To obtain the field produced by Larmor-precessing protons at a test point in the diamond, we need to integrate over the average moment of dipoles for each point in the sample volume. Projecting the field from each sample dipole element onto the NV axis, we obtain

$$\vec{B} \cdot \hat{\mu}_{NV} = \frac{\mu_0 \, \rho_{th}}{4\pi} \int \frac{1}{r^3} \, [3(\hat{r} \cdot \vec{m}(t))(\hat{r} \cdot \hat{\mu}_{NV}) - (\vec{m}(t) \cdot \hat{\mu}_{NV})] dV$$

Here $\rho_{th}$ is the equilibrium density of thermally-polarized protons in the sample, $\hat{r}$ is a unit vector along the direction connecting the test point and the infinitesimal sample volume, $\hat{\mu}_{NV}$ is a unit vector parallel to the NV axis, and $\vec{m}(t)$ is the time-varying magnetic moment vector of the precessing proton spin. We can divide the proton magnetic moment into a direction $\hat{m}$ and magnitude $\gamma_p$, and pull the latter out of the integral such that all dimensions are collected in a prefactor:

$$\vec{B} \cdot \hat{\mu}_{NV} = C \int \frac{1}{r^3} \, [3(\hat{r} \cdot \hat{m}(t))(\hat{r} \cdot \hat{\mu}_{NV}) - (\hat{m}(t) \cdot \hat{\mu}_{NV})] dV$$

The prefactor that determines the approximate scale of the field produced by the sample is

$$C = \frac{\mu_0 \, \gamma_p}{4\pi} \rho_{Tot} \left[ 1 - \exp\left( \frac{-2\gamma_p B_0}{k_B T} \right) \right] \approx \frac{\mu_0 \, \gamma_p}{4\pi} \rho_{Tot} \left[ \frac{2\gamma_p B_0}{k_B T} \right]$$

where $\rho_{Tot}$ is the full density of protons in the sample (taken here to be pure water), $k_B$ is the Boltzmann constant, $T$ is the temperature, and $B_0$ is the magnitude of the static magnetic bias field. Numerically, we obtain

$$C = \frac{\mu_0 \, \gamma_p^2 \rho_{Tot} B_0}{2\pi \, k_B T} = \frac{[4\pi \times 10^{-7} \frac{T \cdot m}{A}][1.41 \times 10^{-26} A \cdot m^2]^2 [6.6 \times 10^{28} m^{-3}][0.0882\, T]}{2\pi \left[ 1.38 \times 10^{-23} \frac{T \cdot A \cdot m^2}{K} \right][300\, K]} = \mathbf{56\ pT}$$

Thus, we expect the magnetic field generated by the protons at the NVs to be on the order of ~50 pT, assuming that the volume integral $G = \int \frac{dV}{r^3} [3(\hat{r} \cdot \hat{m}(t))(\hat{r} \cdot \hat{\mu}_{NV}) - (\hat{m}(t) \cdot \hat{\mu}_{NV})]$ evaluates to something of order ~1 for our sample geometry.

Now we can evaluate the dimensionless integral using the sample geometry defined in Figure S2(a). The z-axis is taken to be perpendicular to the diamond surface. We define $\hat{\mu}_{NV} = (\sqrt{\frac{2}{3}}, 0, \frac{1}{\sqrt{3}})$, as well as two perpendicular unit vectors, $\hat{q}_1 = (\frac{1}{\sqrt{3}}, 0, -\sqrt{\frac{2}{3}})$ and , $\hat{q}_2 = (0,1,0)$ that describe the plane in which the nuclear spins precess when the bias field $\vec{B}_0$ is aligned parallel to the NV axis. Thus, the proton magnetization vector is $\hat{m}(t) = \hat{q}_1 \cos(\omega t) + \hat{q}_2 \sin(\omega t)$. By symmetry, the integral $G$ vanishes for the component of $\hat{m}$ along $\hat{q}_2$, and the component along $\hat{q}_1$ determines the amplitude of the oscillating magnetic field at the sensor. Because $G$ is dimensionless, it is scale-independent, in the sense that for a given shape of the sample volume V, the value of $G$ depends only on the ratio $V^{1/3} / d_{NV}$, where $d_{NV}$ is the depth of the NV test point below the diamond surface. Thus, for a sufficiently large and homogeneous sample volume ($V \gg t^3$, for $t$ the NV layer thickness), every NV center in the sensor feels approximately the same magnetic field from the precessing sample spins.

We evaluate $G$ as a function of $V^{1/3} / d_{NV}$ for hemispherical and cube-shaped volumes, as shown in Figure S2(b). Empirically, we find that the asymptotoic value (as $V^{1/3} / d_{NV} \to \infty$) of $G$ is between 1 and 4. It tends to be largest for sample volumes of high aspect ratio (long in $z$, narrow in $x$ and $y$), which may be of practical importance for sample-holder design in future NV NMR detectors. (For a hemispherical volume, the asymptotic value of $G$ is ~1.4, giving a predicted magnetic signal amplitude of 81 pT.) For all sample volume shapes we have calculated, $G$ reaches half its asymptotic value for $V^{1/3} / d_{NV}$ in the range of 2 – 3. Thus, half of the detected signal in our experiments (with a semi-infinite sample volume) is due to protons within about 25 µm of the sensor, equivalent to a volume of ~30 pL.

Finally, we would like to compare the magnetic noise due to statistically fluctuating sample spin polarization to the mean magnetic signal we just calculated. Following the calculations in reference [12] of the main text, we obtain the variance of the magnetic field fluctuations at a test point of depth $d_{NV}$ below the diamond surface due to a semi-infinite proton sample volume:

$$\Delta B^2 = \frac{\rho_{Tot} \mu_0^2 \gamma_p^2}{96 \pi d_{NV}^2}$$

We want to find the condition on $d_{NV}$ to ensure $\vec{B} \cdot \hat{\mu}_{NV} \geq \sqrt{\Delta B^2}$. Substituting from the expressions above, we find this is equivalent to

$$d_{NV} \geq \left(\frac{k_B T}{\gamma_p B_0 G}\right)^{2/3} \left(\frac{1}{\rho_{Tot}}\right)^{1/3}$$

Numerically, for $B_0$ = 88 mT and $G \sim 1.5$ this evaluates to $d_{NV} > 3$ µm. Thus, the shallowest NV centers in our sensor (in the top 3 µm of the sensing volume) are primarily sensitive to proton spin fluctuations, while the deeper NVs give a signal that is mostly dependent on the thermal proton polarization.

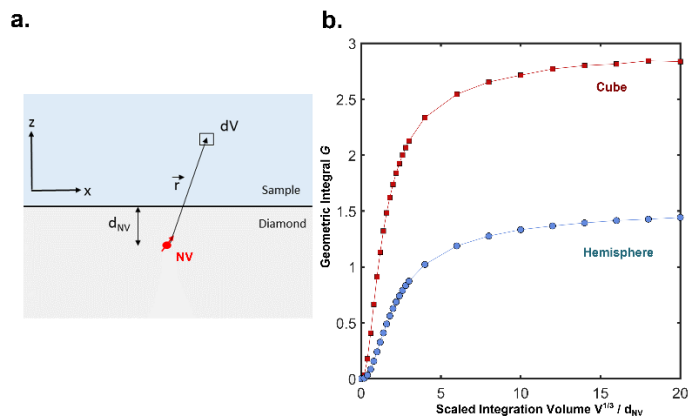

**Figure S2**: Integration of geometric factor G for evaluating magnetic signal strength in the ensemble NV magnetic sensor. **(a)** Schematic of geometry. **(b)** Integrated geometric factor for hemisphere volume (blue circles) or cubic volume (red squares).

## Supplementary Figure 3: Electromagnet Stabilization

The static bias field $B_0$ generated by the electromagnet was actively stabilized using two feedback systems. On short timescales, the field-stabilization relied on a secondary diamond magnetometer (using cw ESR measurements), which was positioned between the electromagnet poles, approximately ~1 cm away from the primary ensemble NV NMR sensor (figure S3a). This secondary diamond magnetometer was optimized for rapid static field measurements, allowing a feedback bandwidth of ~12.5 Hz. To correct slow drifts between the primary ensemble NV NMR sensor and the secondary field-stabilization sensor, we periodically (every 5 minutes) paused the SR protocol and performed pulsed ESR measurements on the primary diamond. Any detected field drift was then used to correct the feedback setpoint of the loop containing the fast (secondary) sensor. The $B_0$ field at the position of the primary sensor, recorded over 2 days, is shown in figure S3b. The magnetic field fluctuations are Gaussian, with standard deviation ~50 nT (figure S3c). Because the measurement precision of the sensors and the current precision of the coil used to correct $B_0$ are both much smaller than the actual $B_0$ fluctuations, the deviation from the setpoint is effectively zero immediately after every feedback adjustment. Assuming linear drift of the magnetic field during each slow (5 minute) feedback interval, the average magnetic field deviation from the setpoint during the interval is approximately half the value recorded at the end of the interval. The real standard deviation of magnetic field fluctuations at the primary ensemble NV NMR sensor is therefore ~25 nT. We therefore expect that these residual B0 fluctuations will limit the observed proton linewidth to approximately $\Gamma \approx 2 \, (2 \ln(2))^{1/2} \times (25 \text{ nT}) \times (42.58 \text{ MHz / T}) = $ **2.5 Hz** FWHM.

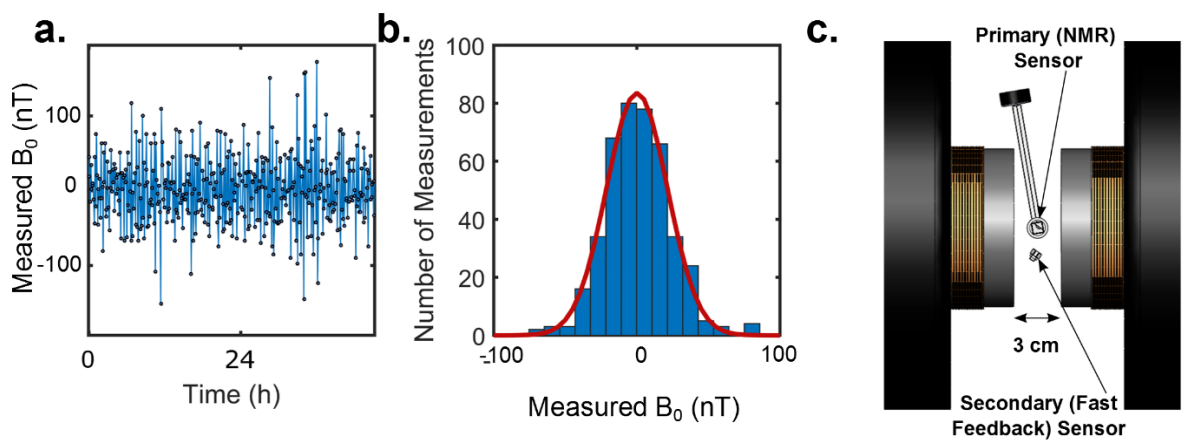

**Figure S3:** Electromagnet Stabilization. **(a)** Time series data of magnetic field deviations, recorded at the primary (NMR) NV ensemble sensor, once every 5 minutes, over 48 hours. **(b)** Histogram of the data in (a), showing a Gaussian distribution of magnetic field deviations with standard deviation 25 nT. **(c)** Schematic of the electromagnet and sensors, drawn to scale. Black coils are the main magnet coils (88 mT); copper coils are the correction coils for fast control of $B_0$.

# Supplementary Table 4: Modified Pulse Sequences to Investigate Sensor-Induced Line Broadening

We experimentally investigated two effects that could potentially lead to broadening of the NMR spectral width due to interaction with the NV sensor:

(i) The microwave field used to drive the NV spins could interact off-resonantly with the protons, resulting in an AC-Zeeman shift of the proton energy levels whenever the microwave field is applied. The approximate magnitude of this effect would be $\Delta f \sim \frac{\Omega_R^2}{\Delta}\left(\frac{\gamma_p}{\gamma_{NV}}\right)^2 \sim 1$ Hz, for $\Omega_R \approx 15$ nominal MHz the NV Rabi frequency, $\Delta \approx 400$ MHz the detuning, $\gamma_p$= 42.58 MHz/T the proton gyromagnetic ratio, and $\gamma_{NV}$= 28.02 GHz/T the NV gyromagnetic ratio. To test for this effect, we decreased $\Omega_R$ by up to a factor of 3, with the duration $\tau$ of each pulse in the NV magnetometry pulse sequence increased proportionally to compensate. Because the expected nuclear phase shift for each pulse is $\Delta\phi = \tau\,\Delta f$, any line broadening should be linearly proportional to $\Omega_R$.

(ii) During the magnetometry pulse sequence, the NV centers are in a superposition of $|m_s=0\rangle$ and $|m_s=-1\rangle$ states, resulting in a net magnetization along the NV quantization axis. This results in a magnetic field gradient outside the diamond, which could produce inhomogeneous broadening of the proton spectra. Numerical calculations (Supplementary Figure 5) suggest this effect should only be relevant for a small fraction of the nuclear sample, and should result in broadening of those spins on the order of ~1 – 2 Hz. To verify this, we used NV magnetometry sequences (i.e. XY8-2, XY8-4, XY8-6) of varying duration $\tau_{seq}$, while keeping the synchronized readout repetition period $\tau_{SR}$ = 24.06 µs fixed. If the gradient produced by the NV magnetization contributed significantly to the proton linewidth, the amount of broadening should be proportional to the fraction of time $\tau_{seq}/\tau_{SR}$ that the NV centers were in the superposition state.

We carried out these experiments on a sample of pure water, which should have FID linewidth < 1 Hz in the absence of field gradients or other perturbations. The results are summarized in the following tables:

| NV Rabi Frequency $\Omega_R$ | Measured FID Linewidth $\Gamma$ |
|---|---|
| 16.6 MHz | 10 ± 1 Hz |
| 8.3 MHz | 8 ± 1 Hz |
| 5.6 MHz | 7 ± 1 Hz |

| Magnetometry Sequence | $\tau_{seq}/\tau_{SR}$ | Measured FID Linewidth $\Gamma$ |
|---|---|---|
| XY8-6 | 0.53 | 9 ± 1 Hz |
| XY8-4 | 0.36 | 8 ± 1 Hz |
| XY8-2 | 0.18 | 10 ± 1 Hz |

Because no significant variation in the measured water SR-FID linewidths was observed in either experiment, we conclude that interaction with the NV ensemble sensor is not the primary source of proton dephasing in our system.

## Supplementary Figure 5: Classical NV Back-Action Calculation

We consider a sensing volume of NV centers at the diamond surface of approximately $(10 \ \mu m)^3$, with polarized density $[NV]_{pol} \approx 0.8 \times 10^{17} cm^{-3}$. (This is equal to one fourth of the estimated total NV density in our sensor, since only NV centers aligned parallel to $B_0$ contribute to sensor back-action.) During the magnetometry sequence, NV centers are in an equal superposition of $|m_s=0\rangle$ and $|m_s=-1\rangle$ states, and thus have an average magnetization of -0.5 $\mu_B$ along the NV axis. The NVs are aligned with the external bias field, so the effective NV magnetization direction is parallel to the proton quantization axis. The NV magnetization will produce a field that is strongest near the diamond surface, and falls off with a length scale on the order of ~1 – 10 μm. This spatial inhomogeneity may broaden the effective resonance width of the protons being sensed. Furthermore, for narrow-band magnetometry with multiple readouts synchronized to an external clock, each NV optical readout repolarizes the NV centers to $|m_s=0\rangle$, temporarily turning off the NV-induced field for several μs and resulting in a (spatially inhomogeneous) phase jump in the proton precession at each readout. We therefore wish to obtain a numerical estimate for the strength of the field gradient produced by the NVs in the proton volume when they are in the superposition state.

The magnetic field at position $\vec{r}$ in the proton sample volume, produced by NV spins of density $\rho_{NV}$, is given by

$$\vec{B}(\vec{r}) = \frac{\mu_0 \rho_{NV}}{4\pi} \int \frac{1}{r^3} [3(\hat{r} \cdot \vec{m}_{NV})\hat{r} - \vec{m}_{NV}] dV$$

where $\vec{m}_{NV} = \frac{\mu_B}{2} \hat{u}_{NV}$ is the NV magnetic moment along the NV axis $\hat{u}_{NV}$, for $\mu_B$ a Bohr magneton, and $\mu_0$ is the permeability of free space. If we consider the projection of this field along the proton quantization axis, which is parallel to $\hat{u}_{NV}$, we obtain:

$$\vec{B} \cdot \hat{u}_{NV} = \left[\frac{\mu_0 \mu_B}{8\pi} \rho_{NV}\right] \int \frac{1}{r^3} [3(\hat{r} \cdot \hat{u}_{NV})^2 - 1] dV$$

The integral contributes a dimensionless geometric factor that depends only on the probe position $\vec{r}$ relative to the NV volume. The dimensional pre-factor sets the scale of magnetic field produced in the proton volume:

$$\left[\frac{\mu_0 \mu_B}{8\pi} \rho_{NV}\right] = \frac{1}{8\pi} [(4\pi \times 10^{-7} \text{ T m/A})(9.3 \times 10^{-24} \text{ J/T})(0.8 \times 10^{23} \text{ m}^{-3})] = 37 \text{ nT}$$

Thus, the expected scale for the magnetic field inhomogeneity due to back-action is on the order of **~37 nT**, or approximately 0.5 ppm of our bias magnetic field $B_0 \sim 88$ mT.

We now compute the geometric integral as a function of space in the proton volume. We take $\hat{u}_{NV} = \sqrt{2/3}\ \hat{x} + \sqrt{1/3}\ \hat{z}$, with the z-axis perpendicular to the diamond surface. We approximate the NV magnetization as two dimensional Gaussian in x and y (with FWHM widths 15 μm and 10 μm respectively) to represent the laser-intensity dependent NV polarization, and a step function in z (such that it is nonzero only between z = - 15 μm and 0 μm) to represent the finite extent of the NV layer below the diamond surface. Under these assumptions, we calculate x-y maps of the geometric integral factor in several planes above the diamond surface (i.e. inside the proton volume) (Figure S6).

Even at a distance of only ~2 μm above the diamond surface, the maximum range of the geometric factor is approximately between ±1, corresponding to a ~±37 nT shift in the $B_0$ field seen by the protons. This might lead to noticeable broadening for high-resolution spectroscopy on the < 1 ppm level, but only from the small fraction of signal protons that are within ~2 μm of the diamond. Over the full detection volume (roughly a hemisphere of radius 20 – 30 μm above the sensor), the mean value of geometric factor is < 0.1, and the expected broadening is on the order of 10 – 100 ppb. Thus, we conclude that while NV back-action might result in minor line-broadening for a small fraction of the signal, it is unlikely to be the dominant factor for linewidths > 1 ppm. (For $B_0$ = 88 mT, 1 ppm of the proton Larmor frequency is about 4 Hz).

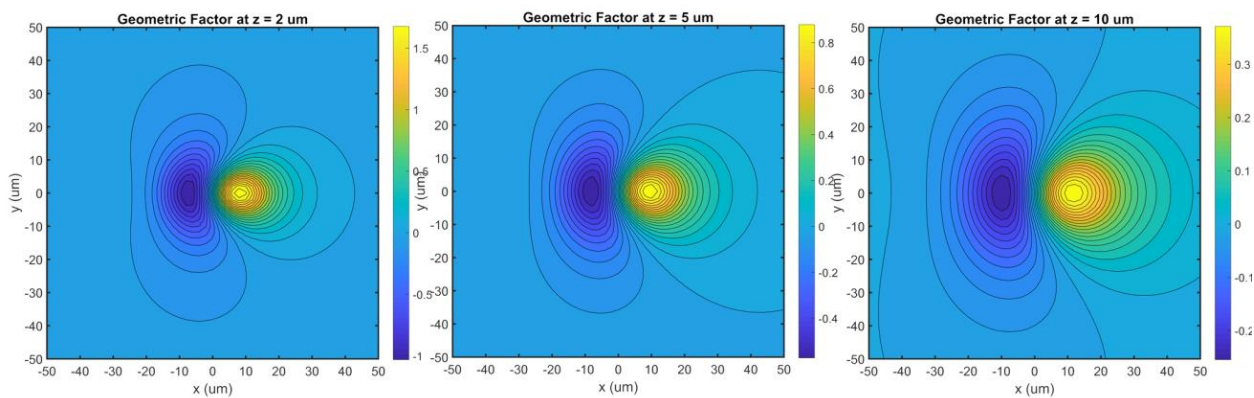

**Figure S5:** Calculated geometric factor as a function of position in the NMR sample. This factor, multiplied by the NV density-dependent scale factor of 37 nT, gives the magnetic field felt by the protons due to the magnetization of the NV centers in the sensor.